\documentstyle[epsfig,12pt]{article}
\def \bea{\begin{eqnarray}}
\def \beq{\begin{equation}}

\def \eea{\end{eqnarray}}
\def \eeq{\end{equation}}
\def \gto{\stackrel{\gamma}{\to}}

\def \ups{\Upsilon}
\begin{document}
\Large
\centerline {\bf Production of the D-wave $b \bar b$ states 
\footnote{Enrico Fermi Institute preprint EFI 01-14, hep-ph/0105273.
Submitted to Physical Review D, Brief Reports.}}
\normalsize
\bigskip
 
\centerline{Stephen Godfrey~\footnote{godfrey@physics.carleton.ca}}
\centerline{\it Ottawa-Carleton Institute for Physics}
\centerline{\it Department of Physics, Carleton University}
\centerline{\it 1125 Colonel By Drive, Ottawa, ON K1S 5B6, Canada}
\smallskip
\centerline{and}
\smallskip
\centerline{Jonathan L. Rosner~\footnote{rosner@hep.uchicago.edu}}
\centerline {\it Enrico Fermi Institute and Department of Physics}
\centerline{\it University of Chicago, 5640 S. Ellis Avenue, Chicago, IL 60637}
\bigskip
 
\begin{quote}

The first and second families of D-wave $b \bar b$ quarkonium states are
expected to have masses near 10.16 and 10.44 GeV/$c^2$.  The accuracy of these
predictions is discussed, and the prospects of two methods for producing these
states in electron-positron collisions are updated.  Direct scans in the
$e^+ e^-$ center-of-mass can give rise to the $^3D_1$ states.  The $1^3D_J$
states have also been searched for in the electromagnetic cascades
from $\ups(3S) \to \gamma \chi'_b \to \gamma \gamma ^3D_J$.  The
sample of $\ups(3S)$ decays required to definitively observe the $1^3D_J$
states is found to be only somewhat greater than the present world's total.

\end{quote}
\bigskip

\noindent
PACS Categories:  14.40.Gx, 13.20.Gd, 13.40.Hq, 12.39.Ki

\bigskip

The bound states of heavy quarks have provided a key
testing ground for quantum chromodynamics (QCD).  Perturbative QCD
describes the short-distance aspects of the interquark force and many
of the decays of the states, while non-perturbative aspects are probed
by the long-distance interaction and by details of fine-structure
and hyperfine splittings \cite{revs}.

The $^3S_1$ states of charmonium ($c \bar c$) and bottomonium ($b \bar b$)
are easily accessible in both hadronic Drell-Yan processes and
electron-positron collisions, since they couple directly to virtual photons.
These states then decay electromagnetically to others with sizeable branching
ratios.  In this way
both the lowest charmonium state, the $\eta_c = c \bar c(^1S_0)$, and many
P-wave $c \bar c$ and $b \bar b$ states have been discovered.  With
knowledge of the masses of the lowest $\chi_c = c \bar c(^3P_J)$ and the
two lowest $\chi_b = b \bar b(^3P_J)$ families, the interquark interaction
can be mapped out to an extent which permits anticipation of other, as
yet unseen, levels.  Foremost among these levels are the $\eta_b = b \bar b
(^1S_0)$ states, for which we have recently suggested observation
strategies \cite{GRetab}, and the D-wave $b \bar b$ levels.  A candidate
for the lowest $^3D_1$ $c \bar c$ level is $\psi(3770)$, which couples
sufficiently strongly to $e^+ e^-$ to be useful as a copious source of
charmed meson pairs.

In this note we review some predictions of $b \bar b(n^3D_J)$ masses for
$n = 1$ and $n = 2$, where $n$ is the level number.
These levels are expected to lie below
the $B \bar B$ flavor threshold and thus to be quite narrow.  A comprehensive
treatment of them was presented in Ref.~\cite{KR}.  We compare the
predictions of that work with others, estimate the likely errors in predictions
of masses, and update the prospects for discovering some of the predicted
levels.  These questions have taken on renewed interest as a result of
plans by the CLEO Collaboration \cite{CLEOrun} to examine some aspects
of $\Upsilon$ spectroscopy, both via significant augmentation of the world's
sample of $\Upsilon(3S)$ decays and via direct scans for $^3D_1$ states
in the 10.13--10.17 and 10.42--10.46 GeV/$c^2$ mass range.

We shall show that most potential models predict a narrow range of
values for the D-wave masses.  The question has been raised of whether
{\it any} potential description is valid for quarkonium \cite{ITEP};
discovery of the D-wave states within the predicted range would not lay
such doubts to rest, but would be one further point in favor of a description
which, at least for the $b \bar b$ system, has been remarkably successful.

\begin{figure}
\centerline{\epsfysize = 4in \epsffile{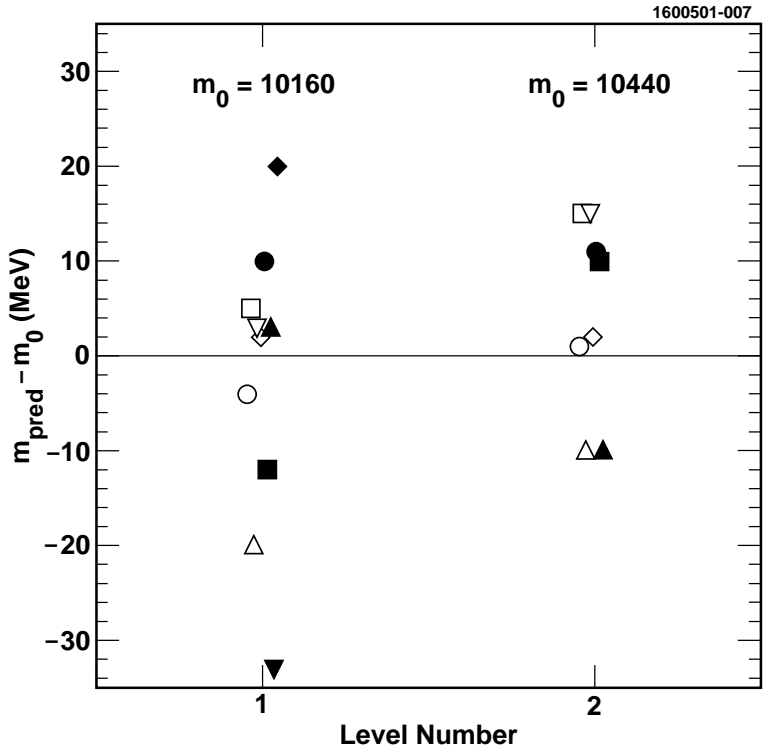}}
\caption{Predictions for the spin-weighted averages of $^{3}D_{J}~b
\overline{b}$ states.  Open circle ($\circ$):  KR \cite{KR} (inverse
scattering); open square ($\Box$):  Cornell \cite{Corn} (QCD-based potential);
open triangle ($\bigtriangleup$): BT \cite{BT} (QCD-based potential;
published masses only quoted to nearest 10 MeV); open inverted triangle
($\bigtriangledown$): Gupta {\it et al.} \cite{GRRe} (QCD-based potential);
open diamond ($\Diamond$): MR \cite{MR} (QCD-based potential);
solid circle: MB \cite{MB} (relativistic corrections; mass of  $^{3}D_{1}$
plotted); solid square: GI \cite{GI} (QCD-based potential, masses calculated
to nearest MeV); solid triangle: Grant {\it et al.} \cite{GRRy} (power-law
potential); solid inverted triangle: EQ \cite{EQ} (QCD-based potential
\cite{BT}, quoted for $n = 1$): solid diamond: lattice \cite{latD}
(quenched approximation with $\beta = 6.0$, quoted for $n = 1$).
\label{fig:Dmas}}
\end{figure}

Some predictions of spin-weighted masses of the $n^3D_J$ $b \bar b$ levels
\cite{KR,Corn,BT,GRRe,MR,MB,GI,GRRy,EQ,latD} are summarized in
Fig.\ \ref{fig:Dmas} \cite{RSG}.
Most potentials give a center-of-gravity of the $1D$ levels within $\pm 10$
MeV/$c^2$ of $10160$ MeV/$c^2$, except for the more recent treatment
of Eichten and Quigg \cite{EQ}, based on the Buchm\"uller-Tye \cite{BT}
potential, which gives a value about 33 MeV/$c^2$
lower.  However, their model also underestimates several other $b \bar b$
masses in comparison with experimental values \cite{PDG}, as shown in Table
\ref{tab:obsm}.  A quenched lattice calculation  \cite{latD} gives a
$1^1D_2$--$1^3S_1$ splitting of 761(20) MeV/$c^2$ for $\beta \equiv 6/g^2_{\rm
QCD} = 6.0$ if the $1P$--$1S$ splitting is used to set the scale. This becomes 
720(30) MeV/$c^2$ if the $2S$-$1S$ splitting is used to set the scale 
(see \cite{latter} in which it is pointed out that this scale is the 
one that makes the bare quark mass used closest to the $b$). 
The result with the $2S$--$1S$ scale is quoted in
Fig.\ \ref{fig:Dmas},
resulting in a $^3D_J$ mass of 10180 MeV if one assumes a small singlet-triplet
splitting in the $1D$ levels.

\begin{table}
\caption{Deviations of predicted masses for $b \bar b$ levels
from observed values, in MeV/$c^2$. \label{tab:obsm}}
\begin{center}
\begin{tabular}{l r r r r r r} \hline
Reference & $\ups(1S)$ & $\chi_{bJ}(1P)$ & $\ups(2S)$ & $\chi_{bJ}(2P)$ &
  $\ups(3S)$ & $\ups(4S)$ \\ \hline
\cite{KR} & (a) & 3 & (a) & $-1$ & (a) & (a) \\
\cite{Corn} & (a) & 23 & (a) & 11 & 3 & 50 \\
\cite{BT} (b) & (a) & $-10$ & 0 & $-10$ & 0 & 40 \\
\cite{GRRe} & 2 & 0 & $-10$ & $-2$ & 0 & (c) \\
\cite{MR} & (a) & 6 & (a) & $-2$ & $-5$ & 27 \\
\cite{MB} & (a) & 23 & (a) &  7  &  0  & 40 \\
\cite{GI} (d) & 5 & $-16$ & $-20$ & $-8$ & $-1$ & 55 \\
\cite{GRRy} (e) & $-5$ & $-9$ & 19 & $-2$ & 4 & $-7$ \\
\cite{EQ} & 4 & $-17$ & $-16$ & $-29$ & $-16$ & 22 \\
\cite{latter} & (a) & $-13 \pm 17$ & (a) & $105 \pm 40$ & $30 \pm 80$
  & (c) \\
Expt.~\cite{PDG} (f) & 9460 & 9900 & 10023 & 10260 & 10355 & 10580 \\ \hline
\end{tabular}
\end{center}
\leftline{(a) Input. (b) Published masses quoted to nearest 10 MeV.
(c) Not quoted.}
\leftline{(d) Numbers in this table are based on masses calculated to 1 MeV,
while}
\leftline{\qquad those published in Ref.\ \cite{GI} were rounded to the nearest
10 MeV.}
\leftline{(e) Ref.~\cite{GRRy} quotes
spin-averaged $nS$ masses; values here are for $^3S_1$ states.}
\leftline{(f) Experimental masses quoted to nearest MeV$/c^2$.}
\end{table}

The difference of scales obtained using different experimental splittings is a
feature of quenched lattice simulations. This difference should disappear once
physical dynamical calculations can be done. In the meantime it is an
additional source of uncertainty. The most recent unquenched calculations are
given in Ref.\ \cite{latunq}, but clear signs of the effect of dynamical quarks
are hard to see.  For S-wave and P-wave lattice predictions, published results
in the quenched approximation are given in Ref.\ \cite{latter}.  These are
quoted in Tables \ref{tab:obsm} and \ref{tab:fs} using the $\beta = 6.0$
results and setting the scale from the $2S$--$1S$ splitting, for consistency
with the results quoted in
Fig.\ \ref{fig:Dmas} \cite{CTHD}. 

The predicted centers-of-gravity of the $2D$ levels range from 10430 to 10455
MeV/$c^2$.  These levels are more likely to be affected by coupling \cite{BE}
to $B^{(*)} \bar B^{(*)}$ systems above the flavor threshold of 10560
MeV/$c^2$.  The fine-structure splittings predicted in various models are
summarized in Table \ref{tab:fs}.

The production of the lowest D-wave $b \bar b$ states is most likely in
cascade transitions from the $\Upsilon(3S)$.  In Ref.\ \cite{KR}, two
means of studying these transitions were proposed.  One can
employ the three-photon inclusive transitions $3S \gto 2P \gto 1D \gto 1P$,
or the four-photon transitions $3S \gto 2P \gto 1D \gto 1P \gto 1S \to
l^+ l^-$, which should have considerably less background.  Either
process suffers from backgrounds due to $3S \gto 2P \gto 2S \gto 1P$ and
$3S \to 1S \pi^0 \pi^0$.  In four-photon processes one can eliminate
events in which two pairs of photons are consistent in mass with two
neutral pions.  However, in such processes the cascade $3S \gto 2P \gto 2S
\gto 1P \gto 1S \to l^+ l^-$ leads to events in which the photons from
$2P \gto 2S$ are easily confused with those from $1D \gto 1P$, while those
from $2S \gto 1P$ are easily confused with those from $2P \gto 1D$.

\begin{table}
\caption{Fine-structure splittings for spin-triplet states predicted in
various models. Shown are deviations from spin-weighted centers-of-gravity,
in MeV/$c^2$. \label{tab:fs}}
\begin{center}
\begin{tabular}{l r r r r r} \hline
Reference   & $J$   & $1P$ & $2P$ & $1D$ & $2D$ \\ \hline
\cite{KR}   & $L-1$ & $-40.4$ (a) & $-28.9$ (a) & $-6.6$ & $-6.3$ \\
            &  $L$  &  $-8.3$ (a) & $-6.3$  (a) & $-0.5$ & $-0.6$ \\
            & $L+1$ &   13.1  (a) &   9.5   (a) &   3.2  &   3.1  \\ \hline
\cite{GRRe} & $L-1$ &   $-32$   &   $-26$   &  $-8$  &  $-8$  \\
            &  $L$  &    $-7$   &    $-6$   &  $-1$  &  $-1$  \\
            & $L+1$ &     10    &     8     &    4   &    4   \\ \hline
\cite{MR}   & $L-1$ &   $-29$   &   $-21$   & $-11$  &  $-9$  \\
            &  $L$  &    $-3$   &    $-2$   &  $-1$  &  $-1$  \\
            & $L+1$ &      8    &     6     &    6   &    5   \\ \hline
\cite{MB}   & $L-1$ &   $-56$   &   $-46$   &   (c)  &   (c)  \\
            &  $L$  &    $-7$   &    $-6$   &   (c)  &   (c)  \\
            & $L+1$ &     15    &     13    &   (c)  &   (c)  \\ \hline
\cite{GI} (b) & $L-1$ & $-37$   &   $-26$   & $-10$  &  $-9$  \\
            &  $L$  &    $-8$   &    $-6$   &  $-1$  &  $-1$  \\
            & $L+1$ &     13    &     9    &    7   &   5    \\ \hline
\cite{EQ}   & $L-1$ &   $-39$   &   $-32$   &  $-7$  &  (c)   \\
            &  $L$  &    $-9$   &    $-7$   &  $-1$  &  (c)   \\
            & $L+1$ &     13    &     11    &   3    &  (c)   \\ \hline
\cite{latter}  & $L-1$ &   $-26$   &    (c)    &   (c)  &  (c)   \\
            &  $L$  &    $-9$   &    (c)    &   (c)  &  (c)   \\
            & $L+1$ &     10    &    (c)    &   (c)  &  (c)   \\ \hline
\end{tabular}
\end{center}
\leftline{(a) Input, based on experimental masses.  (b) Published masses quoted
to nearest 10 MeV;}
\leftline{numbers in this table calculated to 1 MeV.  (c) Not quoted.}
\end{table}

Using the branching ratios predicted in Ref.~\cite{KR} for the electromagnetic
transitions from one $b \bar b$ state to another, and the measured branching
ratio ${\cal B}(\Upsilon(1S) \to e^+ e^-) = (2.38 \pm 0.11)\%$ \cite{PDG}, one
predicts the numbers of $4 \gamma e^+ e^-$ events per $10^6~\Upsilon(3S)$ shown
in Table \ref{tab:nos}.  The total numbers of events proceeding via $1^3D_J$
states (35.1)
and via the $2^3S_1$ state (38.4) are approximately equal, with the dominant
roles played by the transitions $3S \gto 2^3P_1 \gto 1^3D_2 \gto 1^3P_1 \gto
1S$ (20.1 events) and $3S \gto 2^3P_1 \gto 2^3S_1 \gto 1^3P_1 \gto 1S$ (15.9
events).  Equal numbers of $4 \gamma \mu^+ \mu^-$ events are expected if
muons can also be identified.  The initial and final photon energies for these
two sets of transitions are the same, since both sets of transitions proceed
via $^3P_1$ states.  However, the two intermediate pairs of energies are
different:  99 and 261 MeV for the transitions via $1^3D_2$, and 131 and 229
MeV for the transitions via $2^3S_1$.  Thus, it should not be too hard to
distinguish the two processes from one another.

\begin{table}
\caption{Predicted numbers of $4 \gamma e^+ e^-$ events corresponding to
$3S \gto 2P \gto 1D \gto 1P \gto 1S \to e^+ e^-$ or $3S \gto 2P \gto 2S \gto 1P
\gto 1S \to e^+ e^-$ per $10^6~\Upsilon(3S)$
decays.  Numbers following spectroscopic symbols represent photon energies in
MeV, in c.m. of decaying state. \label{tab:nos}}
\begin{center}
\begin{tabular}{c c c c r} \hline
$2^3P_J$ state & Next state & $1^3P_J$ state & $E(\gamma_4)$ & Events \\
\hline
$2^3P_2~(87)$  & $1^3D_3~(107)$ & $1^3P_2~(245)$ & 443 & 7.8 \\
               & $1^3D_2~(112)$ & $1^3P_2~(240)$ & 443 & 0.3 \\
               &                & $1^3P_1~(261)$ & 422 & 2.7 \\
               & $1^3D_1~(119)$ & $1^3P_2~(233)$ & 443 & 0.0 \\
               &                & $1^3P_1~(254)$ & 422 & 0.1 \\
               &                & $1^3P_0~(285)$ & 391 & 0.0 \\ \cline{2-5}
               & $2^3S_1~(242)$ & $1^3P_2~(110)$ & 443 & 4.1 \\
               &                & $1^3P_1~(131)$ & 422 & 8.8 \\
               &                & $1^3P_0~(162)$ & 391 & 0.4 \\ \hline
$2^3P_1~(100)$ & $1^3D_2~(99)$  & $1^3P_2~(240)$ & 443 & 2.5 \\
               &                & $1^3P_1~(261)$ & 422 & 20.1 \\
               & $1^3D_1~(106)$ & $1^3P_2~(233)$ & 443 & 0.1 \\
               &                & $1^3P_1~(254)$ & 422 & 3.3 \\
               &                & $1^3P_0~(285)$ & 391 & 0.4 \\ \cline{2-5}
               & $2^3S_1~(229)$ & $1^3P_2~(110)$ & 443 & 7.5 \\
               &                & $1^3P_1~(131)$ & 422 & 15.9 \\
               &                & $1^3P_0~(162)$ & 391 & 0.7 \\ \hline
$2^3P_0~(123)$ & $1^3D_1~(81)$  & $1^3P_2~(233)$ & 443 & 0.0 \\
               &                & $1^3P_1~(261)$ & 422 & 0.3 \\
               &                & $1^3P_0~(285)$ & 391 & 0.0 \\ \cline{2-5}
               & $2^3S_1~(210)$ & $1^3P_2~(110)$ & 443 & 0.3 \\
               &                & $1^3P_1~(131)$ & 422 & 0.7 \\
               &                & $1^3P_0~(162)$ & 391 & 0.0 \\ \hline
\end{tabular}
\end{center}
\end{table}

The branching ratio predictions of Ref.\ \cite{KR} were performed under the
assumption that the hadronic widths of the D-wave states could be calculated
purely using their color-singlet $b \bar b$ components.  A more up-to-date
calculation of hadronic widths, based on inclusion of color-octet $b \bar b$
contributions \cite{octet}, is probably called for, but is beyond the scope of
the present note.  Even 100\% augmentation of the hadronic widths of the
$1^3D_J$ states would have little effect on their branching ratios for
radiative decays, which are expected to be dominant.  The partial widths
for $2^3D_J$ states to decay non-electromagnetically were not calculated in
Ref.\ \cite{KR} but were expected to be comparable to those for the $1^3D_J$
levels.

We conclude with a discussion of energy scans in $e^+ e^-$ annihilations for
direct production of $b \bar b(^3D_1)$ states.  The results of Table
\ref{tab:fs}
indicate that these states are expected to lie between 6 and 11 MeV/$c^2$ below
the D-wave spin-weighted centers of gravity.  Taking account of
the predictions in
Fig.\ \ref{fig:Dmas}, one then expects the $1^3D_1$ level to lie
between 10.13 and 10.17 GeV/$c^2$, while the $2^3D_1$ level should lie
between 10.42 and 10.46 GeV/$c^2$.  These predictions ignore coupled-channel
distortions due to $B \bar B$ threshold \cite{BE}, as mentioned.

As mentioned in Ref.\ \cite{KR}, present limits on leptonic widths of the
$1^3D_1$ and $2^3D_1$ states \cite{lims} are about a factor of 10 to 15 above
the predicted values \cite{MR} of $\Gamma_{ee}(1^3D_1,2^3D_1) = (1.5,2.7)$ eV.
[Ref.\ \cite{GI} finds $\Gamma_{ee}(1^3D_1) = 1.6$ eV.]
The CUSB Collaboration's search \cite{lims} in the range from 10.34 to
10.52 GeV/$c^2$ sets 90\% c.l. upper limits of $\Gamma_{ee}(2^3D_1) < 40$
eV in this range on the basis of 5 pb$^{-1}$ of integrated luminosity.
Thus, an effective scan of this same range with 15 times the sensitivity
would require at least 1 fb$^{-1}$.  Similar estimates have been obtained
by the CLEO Collaboration \cite{pc}.

Once a $^3D_1$ state has been found in an energy scan, does it have any
distinctive final states?  For the $1^3D_1$ state, the dominant decay, with
60\% branching ratio, was found \cite{KR} to be $\gamma+1^3P_0$, with $E_\gamma
= 285$ MeV.  The $1^3P_0$ is expected to decay 97\% of the time to
hadrons.  However, its hadronic decays do not appear to have the expected
signature of a pair of nearly back-to-back jets \cite{Alam}.
For the $2^3D_1$ state, the corresponding
photon energy for the dominant final state $\gamma+2^3P_0$ is predicted
\cite{KR} to be 202 MeV.  The branching ratio to $\gamma+1^3P_0$, leading to
a 559 MeV photon, is expected to be about a factor of 5 lower.

We thank R. S. Galik and T. Skwarnicki for asking questions which led to this
investigation, and for extensive discussions.  We are grateful to C. T. H.
Davies for communications with regard to lattice gauge theory results.
This work was supported in part by the United
States Department of Energy through Grant No.\ DE FG02 90ER40560
and the Natural Sciences and Engineering Research Council of Canada.

\def \ajp#1#2#3{Am.\ J. Phys.\ {\bf#1}, #2 (#3)}
\def \apny#1#2#3{Ann.\ Phys.\ (N.Y.) {\bf#1}, #2 (#3)}
\def \app#1#2#3{Acta Phys.\ Polonica {\bf#1}, #2 (#3)}
\def \arnps#1#2#3{Ann.\ Rev.\ Nucl.\ Part.\ Sci.\ {\bf#1}, #2 (#3)}
\def \art{and references therein}
\def \cmts#1#2#3{Comments on Nucl.\ Part.\ Phys.\ {\bf#1}, #2 (#3)}
\def \cn{Collaboration}
\def \cp89{{\it CP Violation,} edited by C. Jarlskog (World Scientific,
Singapore, 1989)}
\def \efi{Enrico Fermi Institute Report No.\ }
\def \epjc#1#2#3{Eur.\ Phys.\ J. C {\bf#1}, #2 (#3)}
\def \f79{{\it Proceedings of the 1979 International Symposium on Lepton and
Photon Interactions at High Energies,} Fermilab, August 23-29, 1979, ed. by
T. B. W. Kirk and H. D. I. Abarbanel (Fermi National Accelerator Laboratory,
Batavia, IL, 1979}
\def \hb87{{\it Proceeding of the 1987 International Symposium on Lepton and
Photon Interactions at High Energies,} Hamburg, 1987, ed. by W. Bartel
and R. R\"uckl (Nucl.\ Phys.\ B, Proc.\ Suppl., vol.\ 3) (North-Holland,
Amsterdam, 1988)}
\def \ib{{\it ibid.}~}
\def \ibj#1#2#3{~{\bf#1}, #2 (#3)}
\def \ichep72{{\it Proceedings of the XVI International Conference on High
Energy Physics}, Chicago and Batavia, Illinois, Sept. 6 -- 13, 1972,
edited by J. D. Jackson, A. Roberts, and R. Donaldson (Fermilab, Batavia,
IL, 1972)}
\def \ijmpa#1#2#3{Int.\ J.\ Mod.\ Phys.\ A {\bf#1}, #2 (#3)}
\def \ite{{\it et al.}}
\def \jhep#1#2#3{JHEP {\bf#1}, #2 (#3)}
\def \jpb#1#2#3{J.\ Phys.\ B {\bf#1}, #2 (#3)}
\def \lg{{\it Proceedings of the XIXth International Symposium on
Lepton and Photon Interactions,} Stanford, California, August 9--14 1999,
edited by J. Jaros and M. Peskin (World Scientific, Singapore, 2000)}
\def \lkl87{{\it Selected Topics in Electroweak Interactions} (Proceedings of
the Second Lake Louise Institute on New Frontiers in Particle Physics, 15 --
21 February, 1987), edited by J. M. Cameron \ite~(World Scientific, Singapore,
1987)}
\def \kdvs#1#2#3{{Kong.\ Danske Vid.\ Selsk., Matt-fys.\ Medd.} {\bf #1},
No.\ #2 (#3)}
\def \ky85{{\it Proceedings of the International Symposium on Lepton and
Photon Interactions at High Energy,} Kyoto, Aug.~19-24, 1985, edited by M.
Konuma and K. Takahashi (Kyoto Univ., Kyoto, 1985)}
\def \mpla#1#2#3{Mod.\ Phys.\ Lett.\ A {\bf#1}, #2 (#3)}
\def \nat#1#2#3{Nature {\bf#1}, #2 (#3)}
\def \nc#1#2#3{Nuovo Cim.\ {\bf#1}, #2 (#3)}
\def \nima#1#2#3{Nucl.\ Instr.\ Meth. A {\bf#1}, #2 (#3)}
\def \np#1#2#3{Nucl.\ Phys.\ {\bf#1}, #2 (#3)}
\def \npbps#1#2#3{Nucl.\ Phys.\ B Proc.\ Suppl.\ {\bf#1}, #2 (#3)}
\def \os{XXX International Conference on High Energy Physics, Osaka, Japan,
July 27 -- August 2, 2000}
\def \PDG{Particle Data Group, D. E. Groom \ite, \epjc{15}{1}{2000}}
\def \pisma#1#2#3#4{Pis'ma Zh.\ Eksp.\ Teor.\ Fiz.\ {\bf#1}, #2 (#3) [JETP
Lett.\ {\bf#1}, #4 (#3)]}
\def \pl#1#2#3{Phys.\ Lett.\ {\bf#1}, #2 (#3)}
\def \pla#1#2#3{Phys.\ Lett.\ A {\bf#1}, #2 (#3)}
\def \plb#1#2#3{Phys.\ Lett.\ B {\bf#1}, #2 (#3)}
\def \pr#1#2#3{Phys.\ Rev.\ {\bf#1}, #2 (#3)}
\def \prc#1#2#3{Phys.\ Rev.\ C {\bf#1}, #2 (#3)}
\def \prd#1#2#3{Phys.\ Rev.\ D {\bf#1}, #2 (#3)}
\def \prl#1#2#3{Phys.\ Rev.\ Lett.\ {\bf#1}, #2 (#3)}
\def \prp#1#2#3{Phys.\ Rep.\ {\bf#1}, #2 (#3)}
\def \ptp#1#2#3{Prog.\ Theor.\ Phys.\ {\bf#1}, #2 (#3)}
\def \rmp#1#2#3{Rev.\ Mod.\ Phys.\ {\bf#1}, #2 (#3)}
\def \rp#1{~~~~~\ldots\ldots{\rm rp~}{#1}~~~~~}
\def \rpp#1#2#3{Rep.\ Prog.\ Phys.\ {\bf#1}, #2 (#3)}
\def \sing{{\it Proceedings of the 25th International Conference on High Energy
Physics, Singapore, Aug. 2--8, 1990}, edited by. K. K. Phua and Y. Yamaguchi
(Southeast Asia Physics Association, 1991)}
\def \slc87{{\it Proceedings of the Salt Lake City Meeting} (Division of
Particles and Fields, American Physical Society, Salt Lake City, Utah, 1987),
ed. by C. DeTar and J. S. Ball (World Scientific, Singapore, 1987)}
\def \slac89{{\it Proceedings of the XIVth International Symposium on
Lepton and Photon Interactions,} Stanford, California, 1989, edited by M.
Riordan (World Scientific, Singapore, 1990)}
\def \smass82{{\it Proceedings of the 1982 DPF Summer Study on Elementary
Particle Physics and Future Facilities}, Snowmass, Colorado, edited by R.
Donaldson, R. Gustafson, and F. Paige (World Scientific, Singapore, 1982)}
\def \smass90{{\it Research Directions for the Decade} (Proceedings of the
1990 Summer Study on High Energy Physics, June 25--July 13, Snowmass, Colorado),
edited by E. L. Berger (World Scientific, Singapore, 1992)}
\def \tasi{{\it Testing the Standard Model} (Proceedings of the 1990
Theoretical Advanced Study Institute in Elementary Particle Physics, Boulder,
Colorado, 3--27 June, 1990), edited by M. Cveti\v{c} and P. Langacker
(World Scientific, Singapore, 1991)}
\def \yaf#1#2#3#4{Yad.\ Fiz.\ {\bf#1}, #2 (#3) [Sov.\ J.\ Nucl.\ Phys.\
{\bf #1}, #4 (#3)]}
\def \zhetf#1#2#3#4#5#6{Zh.\ Eksp.\ Teor.\ Fiz.\ {\bf #1}, #2 (#3) [Sov.\
Phys.\ - JETP {\bf #4}, #5 (#6)]}
\def \zpc#1#2#3{Zeit.\ Phys.\ C {\bf#1}, #2 (#3)}
\def \zpd#1#2#3{Zeit.\ Phys.\ D {\bf#1}, #2 (#3)}

\end{document}